\documentclass[twocolumn,aps,prl,superscriptaddress,longbibliography]{revtex4-2}
\usepackage{multibib}
\usepackage{graphicx}
\usepackage{amsmath}
\usepackage{amssymb}
\usepackage{ulem}
\usepackage{mathrsfs}
\usepackage{braket}
\usepackage{bm}
\usepackage[colorlinks=true,linkcolor=blue,citecolor=blue,urlcolor=blue]{hyperref}
\usepackage{subfigure}
\usepackage{graphicx}
\usepackage{physics}
\usepackage{ulem}

\begin{document}
	\title{Enantiospecific Two-Photon Electric-Dipole Selection Rule of Chiral Molecules}
	
	\author{Fen Zou}
	
	\affiliation{Center for Theoretical Physics \& School of Physics and Optoelectronic Engineering, Hainan University, Haikou 570228, China}
	
	\author{Yong Li}
	\email{yongli@hainanu.edu.cn}
	
	\affiliation{Center for Theoretical Physics \& School of Physics and Optoelectronic Engineering, Hainan University, Haikou 570228, China}
	
	\author{Peng Zhang}
	\email{pengzhang@ruc.edu.cn}
	
	\affiliation{School of Physics, Renmin University of China, Beijing, 100872,
		China}
	
	\affiliation{Key Laboratory of Quantum State Construction and Manipulation (Ministry of Education), Renmin University of China, Beijing, 100872, China}
	
	\date{\today}
	
	\begin{abstract}
	Distinguishing between enantiomers is crucial in the study of chiral molecules in chemistry and pharmacology. Many optical approaches rely on enantiospecific cyclic electric-dipole transitions induced by three microwave or laser beams.  However, these approaches impose stringent requirements, including phase locking, three-photon resonance, and precise control over beam intensities and operation times, which enhance the complexity and restrict the applicability. 
	In this letter, we present a novel optical method that {\it eliminates these constraints entirely.}
	Specifically, we demonstrate that in the presence of a static electric field, there is a selection rule for two-photon electric-dipole transitions that differs between enantiomers. This distinction arises because the static electric field breaks the symmetry associated with the combined action of a specific rotation and time-reversal transformation.	Leveraging the enantiospecific two-photon selection rule, one can selectively excite a desired enantiomer using two beams, without the need for phase locking, resonance condition, and the precise control of their  intensities and operation times. Our method significantly enhances the feasibility and applicability of optical approaches  for enantiomer differentiation. 
		
	\end{abstract}
	
	\maketitle
	
	{\it Introduction.---}Chiral molecules are those that have a specific configuration and cannot be superimposed on their mirror images, which are known as enantiomers. The enantiomers with opposite handedness exhibit significant differences in biological activity and chemical behavior~\cite{Mezey1991}, but have nearly identical physical properties.  
	Efficient approaches of enantiomer differentiation, like
enantiodetection~\cite{Patterson2013nature,Patterson2013,Patterson2014,Lobsiger2015,Shubert2016,Yachmenev2016,Ye2019De,Ye2021,Cai2022,Chen2022,Kang2023,Ye2023,Ke2023}, enantioseparation~\cite{Li2007,Li2010,Eilam2013,Liu2021}, and enantiospecific state transfer~\cite{Kral2001,Kra2003,Li2008,Jia2010,Leibscher2019,Vitanov2019,Ye2019,Torosov2020,TorosovBT2020,Wu2019,Wu2020,Guo2022,Leibscher2022,Liu2022,Cheng2023,Zou2024,Eibenberger2017,Perez2017,Perez2018,Lee2022,Sun2023}, are crucial in chemical, biological, and pharmaceutical research~\cite{Eriksson2001,Teo2004}.
	

	Traditional optical approaches for distinguishing enantiomers (e.g., various circular dichroisms~\cite{Stephens1985,Beaulieu2018} and Raman optical activity~\cite{YananHe2011,Fanood2015}) rely on the weak magnetic-dipole interaction or electric-quadrupole interaction between molecules and microwave (or laser) fields, necessitating high-density samples for detectable signals~\cite{Patterson2013nature}.
	Over the past two decades, numerous optical approaches based on electric-dipole interactions—which are significantly stronger than magnetic-dipole interaction or electric-quadrupole interaction—have been proposed and experimentally demonstrated~\cite{Patterson2013nature,Patterson2013,Patterson2014,Lobsiger2015,Shubert2016,Yachmenev2016,Ye2019De,Ye2021,Cai2022,Chen2022,Kang2023,Ye2023,Ke2023,Li2007,Li2010,Eilam2013,Liu2021,Kral2001,Kra2003,Li2008,Jia2010,Leibscher2019,Vitanov2019,Ye2019,Torosov2020,TorosovBT2020,Wu2019,Wu2020,Guo2022,Leibscher2022,Liu2022,Cheng2023,Zou2024,Eibenberger2017,Perez2017,Perez2018,Lee2022,Sun2023,liang:hal-05039436}.
	These approaches exploit cyclic electric-dipole transitions between three molecular rovibrational states, driven by three microwave or laser beams.
However, these approaches require phase locking, three-photon resonance, and precise control over their intensities and/or operation times. These requirements considerably introduce additional complexity and limit the feasibility.
	In particular,
	phase locking is far more complex for laser beams than for microwaves, restricting experimental demonstrations to microwave setups.
	Consequently, the energy gaps between the states involved in the cyclic transition are of the order of the microwave frequencies, much lower than the room-temperature thermal fluctuations. 
	Thus, to mitigate the adverse effects of thermal fluctuations,
	these experiments must either be conducted at ultralow temperatures on the order of kelvins~\cite{Patterson2013nature,Eibenberger2017,Lee2022,Sun2023,Patterson2013} or necessitate the use of additional depletion techniques~\cite{Lee2022}.
	
	In this letter, we propose a novel optical method for enantiomer differentiation, which eliminates the necessity for phase locking, resonance condition, and precise control of intensities and operation times of the microwave or laser beams.
	Our approach  utilizes two microwave or laser beams, as well as a static electric field (E-field). Previous researchers have developed approaches using a weak static electric field and microwave or laser beams to prepare an achiral molecule in a chiral vibrational state~\cite{Tikhonov2022,Leibscher2024}, or to detect the enantiomeric excess in a mixture of chiral molecules with opposite handedness~\cite{Patterson2013nature}. The latter  still requires the phase locking and precise control of the intensities of the beams. Here we show that in the presence of a static E-field, the selection rules for certain two-photon cascade electric-dipole transitions between rovibrational states of chiral molecules become enantiospecific, due to symmetry changes induced by the static  E-field.
Specifically, these  transitions are driven by two microwave or laser beams linearly polarized in a plane perpendicular to the static E-field. The enantiospecific two-photon selection rule (TPSR) refers to the fact that, when the angle between the polarization directions of these two beams reaches a specific value, the transition is forbidden for one enantiomer but allowed for the other one. 
Utilizing this enantiospecific TPSR, one can realize enantiospecific transition to distinguish between the enantiomers. Notably, the enantiospecific TPSR is independent of the phases, detunings, intensities, and operation times of the beams. Therefore, resonance conditions and precise control or locking of these parameters is not required for enantiospecific transition.

		{\it Two-Photon Selection Rule (TPSR).---}We consider a left- (right-) handed chiral molecule \( L \) (\( R \)) in a static E-field with strength \( \bm{E} = E_0 \mathbf{e}_z \) (\( E_0 > 0 \)). Here \( \mathbf{e}_j \) (\( j = x, y, z \)) is the unit vector along the \( j \)-axis of the lab frame. In this work, we assume $E_0$ is weak, such that only the $\bm{E} $--induced coupling between rotational states within the same electronic and vibrational levels needs to be considered. For molecules with dipole moment being on the order of Debye and energy gap between vibrational levels being on the order of $\hbar(2\pi)10^{11}$Hz or larger, this condition can be safely satisfied when 
		\( E_0 \) is on the order of 10 kV/cm or lower. 
		
		\begin{figure}[tbp]
			\includegraphics[width=1.0\columnwidth]{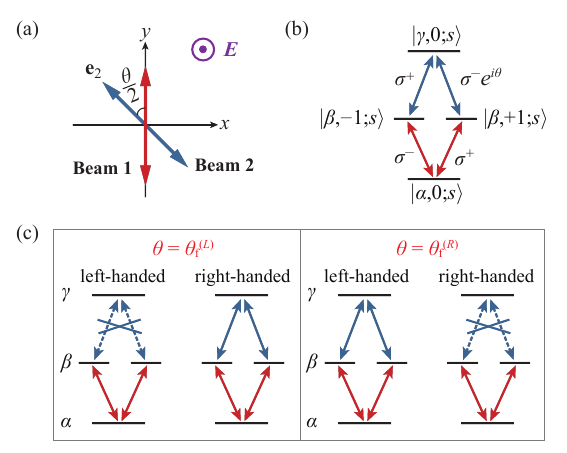}
			\caption{{\bf (a):} The directions of the static E-field $\bm{E}$ (purple) and the polarizations of beams 1 (red) and 2 (blue).  {\bf (b):} The rovibrational levels and the circularly-polarized components of beams 1 (red) and 2 (blue) involved in the cascade transition of Eq.~(\ref{tp}). {\bf (c):} Enantiospecific  transitions: when $\theta=\theta_{\rm f}^{(L/R)}$, the transition is allowed for the right-/left-handed enantiomer, but
			forbidden for the left-/right-handed one.
		 (b) and (c) present the schematics for  $\epsilon_{\alpha, 0}<\epsilon_{\beta,\pm 1}<\epsilon_{\gamma, 0}$.  The ones for $\epsilon_{\alpha, 0}<\epsilon_{\gamma, 0}<\epsilon_{\beta,\pm 1}$ are given in Fig.~\ref{fig4}.}
			\label{fig0}
		\end{figure} 
		
		We denote the Hamiltonian of the rovibrational states of molecule $s$ ($s=L,R$) as $\hat H_0^{(s)}(E_0)$. Since the $z$-component of the total angular momentum of all the  nuclei and electrons in the molecule ($\hat J_z$) is conserved, the eigen-states and eigen-energies of $\hat H_0^{(s)}(E_0)$ can be denoted as $\{|\xi,\!M;\!s\rangle\}$ and $\{\epsilon_{\xi, M}\}$, respectively~\cite{SM}. Here $M=0,\pm 1, \pm 2,...$ is the quantum number of $\hat J_z$. Moreover, $\xi = 1, 2, 3, \dots$ denotes the energy levels for a fixed $M$, and follows the sequence $\epsilon_{1, M} < \epsilon_{2, M} < \epsilon_{3, M} <...$.
		Note that both $|\xi,\!M;\!s\rangle$ and $\epsilon_{\xi, M}$ {\it depend on} the strength $E_0$ of the static E-field. Moreover,  $\epsilon_{\xi, M}$ is independent of the handedness  $s$, and satisfies $\epsilon_{\xi, M}=\epsilon_{\xi, -M}$, which is due to the time-reversal symmetry. Here we neglect the tiny parity-violating energy difference due to the fundamental weak force~\cite{Quack2008}.

		Furthermore, two microwave or laser beams, labeled 1 and 2, are applied to the molecule \( s \) (\( s = L, R \)). Both beams are linearly polarized in the \( x \)-\( y \) plane. Without loss of generality, we assume that beam 1 is polarized along \( \mathbf{e}_y \), while beam 2 is polarized along a direction  \( \mathbf{e}_2 = \sin{(\theta/2)} \mathbf{e}_x - \cos{(\theta/2)} \mathbf{e}_y \) (\( \theta \in (-\pi, \pi] \)) [Fig.~\ref{fig0}(a)]. Additionally, beam 1 is in near-resonance with the transition between a rovibrational state \( |\alpha, 0; s \rangle \) and two degenerate ones \( |\beta, \pm 1; s \rangle \), while beam 2 is near-resonance with the transition between \( |\beta, \pm 1; s \rangle \) and another rovibrational state \( |\gamma, 0; s \rangle \). Therefore, the two beams can induce the two-photon cascade transition:
		\begin{eqnarray}
			|\alpha, 0; s \rangle \rightarrow |\beta, \pm 1; s \rangle \rightarrow |\gamma, 0; s \rangle,\label{tp}
		\end{eqnarray}
		and its inverse process,
 for either  $\epsilon_{\alpha, 0}<\epsilon_{\beta,\pm 1}<\epsilon_{\gamma, 0}$   or $\epsilon_{\alpha, 0}<\epsilon_{\gamma, 0}<\epsilon_{\beta,\pm 1}$.

		The above two-photon transitions  follow a polarization selection rule, which can be derived as follows. Since
		\begin{eqnarray}
			\mathbf{e}_y\propto (\mathbf{e}_++\mathbf{e}_-),\hspace{0.5cm}
			{\mathbf{e}}_2\propto (\mathbf{e}_++e^{i\theta}\mathbf{e}_-) \label{eyd}
		\end{eqnarray}
		with  $\mathbf{e}_\pm\equiv\mp(\mathbf{e}_x\pm i\mathbf{e}_y)/\sqrt{2}$, the beam 1(2) can be decomposed as a superposition of the circularly polarized  beams
		$\sigma^{\pm}$ , which are polarized along
		$\mathbf{e}_{\pm}$ , with a relative phase $0$($\theta$). Using this result and the selection rules for the electric-dipole couplings induced by circularly polarized electromagnetic waves [Fig.~\ref{fig0}(b) and Fig.~\ref{fig4}(a)],  
		we find that 
		beam 1 couples  $|\alpha,\!0;\!s\rangle$ only to the dressed state 
		\begin{eqnarray}
			|\psi;s\rangle= a_{-}^{(s)}|\beta,\! -1;\!s\rangle+a_{+}^{(s)}|\beta,\! +1;\!s\rangle,\label{psis}
		\end{eqnarray}
		where
		\begin{eqnarray}
			a_{\pm }^{(s)}=\langle \beta,\!\pm 1;\!s|\hat {\bm d}\cdot{\mathbf{e}}_\pm| \alpha,\!0;\!s\rangle,
			\label{apm0}
		\end{eqnarray}
		with $\hat  {\bm d}$ being the molecular electric-dipole operator. 
		Thus,  the cascade transition (\ref{tp}) essentially involves only one intermediate state
		(i.e., $|\psi;s\rangle$).
		Additionally, 
		the  matrix element of the beam-2--induced electric-dipole coupling 
		between $|\psi;s\rangle$ and $|\gamma,\!0;s\rangle$ satisfies
		\begin{eqnarray}
			\langle \gamma,\!0 ; s|\hat {\bm d}\cdot{\mathbf{e}}_2| \psi; s\rangle\propto 
			\bigg(a_{-}^{(s)}b_{+}^{(s)}+a_{+}^{(s)}b_{-}^{(s)}e^{i\theta} \bigg),\label{prop}
		\end{eqnarray}
		where  
		\begin{eqnarray}
			b_{\pm }^{(s)}=\langle \gamma;0;\!s|\hat {\bm d}\cdot{\mathbf{e}}_\pm| \beta,\!\mp 1;\!s\rangle.\label{bpm0}
		\end{eqnarray}
		Moreover, due to the time-reversal symmetry, we have
		$|a_{+}^{(s)}|=|a_{-}^{(s)}|$ and $|b_{+}^{(s)}|=|b_{-}^{(s)}|$. This fact and Eq.~(\ref{prop}) yield that 
		the matrix element $\langle \gamma,\!0 ; s|\hat {\bm d}\cdot{\mathbf{e}}_2| \psi; s\rangle$ becomes zero, and thus the cascade transition (\ref{tp}) is forbidden for the molecule $s$ ($s=L,R$), under the condition
		$\theta=\theta_{\rm f}^{(s)}(E_0)$, with $\theta_{\rm f}^{(s)}(E_0)$ being defined as
		\begin{eqnarray}
			\theta_{\rm f}^{(s)}(E_0):=
			{\rm arg}\left[-a_{+}^{(s)\ast}
			b_{-}^{(s)\ast}
			b_{+}^{(s)}
			a_{-}^{(s)}\right].\label{tast}
		\end{eqnarray}
		Here ${\rm arg}[z]\in (-\pi,\pi]$ for $ z\in {\mathbb C}$. We call $\theta_{\rm f}^{(L/R)}(E_0)$ as the forbidden polarization angle of molecule $L/R$.
		
		Now we have obtained the TPSR of the cascade transition in Eq.~(\ref{tp}). Specifically, this transition is forbidden for the left-handed or right-handed chiral molecule, when $\theta=\theta_{\rm f}^{(L)}(E_0)$ or $\theta=\theta_{\rm f}^{(R)}(E_0)$, respectively. This TPSR and the expressions  of $\theta_{\rm f}^{(s)}(E_0)$, $a_{\pm }^{(s)}$ and $b_{\pm }^{(s)}$ [i.e., Eqs.~(\ref{apm0}-\ref{tast})] are applicable for both 
		$\epsilon_{\alpha, 0}<\epsilon_{\beta,\pm 1}<\epsilon_{\gamma, 0}$  and $\epsilon_{\alpha, 0}<\epsilon_{\gamma, 0}<\epsilon_{\beta,\pm 1}$.

		We emphasis that the forbidden polarization angles $\theta_{\rm f}^{(L,R)}(E_0)$ are independent of the  phases, detunings, intensities, and operation times of beams 1 and 2. This is because $\theta_{\rm f}^{(L,R)}(E_0)$ are determined by the factors $a_{\pm }^{(L,R)}$ and $b_{\pm }^{(L,R)}$, which, according to Eqs.~(\ref{apm0}, \ref{bpm0}), are independent of the aforementioned beam parameters.

		
		%
		%

		{\it Enantiospecific TPSR.---}Now there is a natural question: 
		
		\vspace{0.3cm}
		
		\noindent {\it Are the forbidden polarization angles $\theta_{\rm f}^{(L)}(E_0)$ and $\theta_{\rm f}^{(R)}(E_0)$ of the left- and right-handed molecules the same or not?} 
		
		\vspace{0.3cm}
		\noindent 
		It is clear that  \( \theta_{\rm f}^{(L)} = \theta_{\rm f}^{(R)} \) in the absence of the static E-field. However, when the static E-field is applied (i.e., when \( E_0 \neq 0 \)), we generally find that \( \theta_{\rm f}^{(L)} \neq \theta_{\rm f}^{(R)} \). In other words, the TPSR becomes enantiospecific.

		Before providing an  explanation for this conclusion, we first illustrate it using the results for a typical chiral molecule. We calculate the forbidden polarization angles $\theta_{\rm f}^{(L,R)}(E_0)$  for a left- or right-handed 1,2-propanediol molecule in the static E-field with $E_0 \leq 20\, \text{kV/cm}$,
		with various specific choices of states
		$|\alpha,0,s\rangle$, $|\beta,\pm 1,s\rangle$, and $|\gamma,0,s\rangle$ ($s=L,R$), and present the  most representative results   in Figs.~\ref{result}(a, b).
		 The details of the calculation and more results are given in the Supplementary Material (SM)~\cite{SM}.
		These results indicate that, in general, $\theta_{\rm f}^{(L)}(E_0) \neq \theta_{\rm f}^{(R)}(E_0)$ for a non-zero $E_0$. Additionally, the figures suggest that $\theta_{\rm f}^{(L)}(E_0) =- \theta_{\rm f}^{(R)}(E_0)$, as will be demonstrated later.
		
		%
		
		We further define the forbidden polarization  directions 
		$\mathbf e_{\rm 2f}^{(s)}$ ($s=L,R$)
		as
		$\mathbf e_{\rm 2f}^{(s)}(E_0):=\sin[{\theta_{\rm f}^{(s)}(E_0)/2}] \mathbf{e}_x - \cos[{\theta_{\rm f}^{(s)}(E_0)/2}] \mathbf{e}_y$. Thus, the TPSR can be re-expressed as: the cascade transition (\ref{tp}) is forbidden for molecule $s$ when 
		$\mathbf e_2=\mathbf e_{\rm 2f}^{(s)}$. 
		Therefore, the degree of enantiospecificity of the TPSR   can be defined as
		\begin{eqnarray}
			D&:=&\big|\mathbf e_{\rm 2f}^{(L)}\times \mathbf e_{\rm 2f}^{(R)}\big|^2
			=\sin^{2}\left[\frac{1}{2}\left(\theta^{(L)}_{\rm f}-\theta^{(R)}_{\rm f}\right)\right], 		
			\end{eqnarray}
		which satisfies  $D\in[0,1]$.
		The TPSR is enantiospecific ($\mathbf e_{\rm 2f}^{(L)}\neq
		\mathbf e_{\rm 2f}^{(R)}
		$) if $D \neq 0$.
		In particular, in the case with $D=1$
		(i.e., $\mathbf e_{\rm 2f}^{(L)}\perp \mathbf e_{\rm 2f}^{(R)}$),
		when the  cascade transition is forbidden  for the left-handed enantiomer  (i.e., $\langle \gamma,\!0 ; L|\hat {\bm d}\cdot{\mathbf{e}}_2| \psi; L\rangle=0$), 
		the absolute value of  the matrix element $\langle \gamma,\!0 ; R|\hat {\bm d}\cdot{\mathbf{e}}_2| \psi; R\rangle$ of the right-handed enantiomer  just reaches its maximum value, and vice versa. In this case the 
		TPSR  is enantiospecific  to the maximum extent. 
		
		\begin{figure}[tbp]
			\includegraphics[width=0.85\columnwidth]{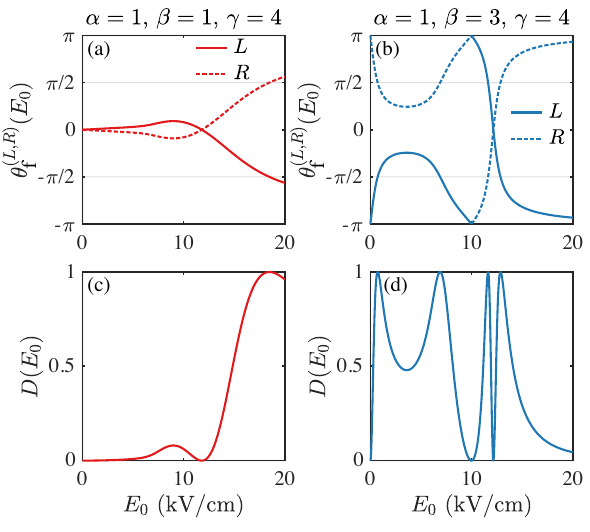}
			\caption{{\bf (a, b):} The forbidden polarization angle $\theta_{\rm f}^{(L,R)}(E_0)$. {\bf (c, d):} The  degree  $D(E_{0})$  of enantiospecificity
of the TPSR. Here we show the results for  transitions of Eq.~(\ref{tp}) for 1,2-propanediol molecules, with relevant quantum numbers being $(\alpha=1;\beta=1;\gamma=4)$ (a, c), and $(\alpha=1;\beta=3;\gamma=4)$ (b, d).}
			\label{result}
		\end{figure}

		In Figs.~\ref{result}(c, d), we illustrate the value of \( D \) for 1,2-propanediol molecule as a function of the static E-field strength \( E_0 \). These results, along with those presented in the SM \cite{SM} for additional cases, clearly demonstrate that when \( E_0 \) reaches 1-20 kV/cm, situations usually arise where \( D \) becomes  comparable to its maximum value of 1, indicating that the TPSR is highly enantiospecific.

	   Now we present the origin of the enantiospecific TPSR through a symmetry analysis.  	 
	   For convenience, we denote $x,y,z$ as the spatial coordinates of the particles (atomic nucleus and electrons) of each molecule, in the center-of-mass  (CoM) frame of this molecule.  Specifically, $x:=\{x_1,x_2,...\}$, where $x_{i}$ ($i=1,2,...$) is the $x$-coordinate the particle $i$, and $y$ and $z$ are defined similarly. Note that the origin of the CoM frame is located at the CoM position of the molecule, and the axes of this frame are parallel to those of the lab frame.
	   
We perform our analysis in the Hilbert space $\mathscr H$ spanned by both sets of basis of $\{|\xi,\! M;\!L\rangle\}$ and $\{|\xi,\! M;\!R\rangle\}$.
Every state $|\Psi\rangle\in \mathscr H$  is a state of the relative motion of these particles, and can be described by a wave function $\langle x,y,z|\Psi\rangle$.

Here we introduce the transformations:
		\begin{itemize}
			\item[(i)]  ${\hat {\cal P}}$: $(x,y,z)\rightarrow(-x,-y,-z)$ (spatial inversion).
			\item[(ii)] ${\hat {\cal C}}_{2x}$: $(x,y,z)\rightarrow(x,-y,-z)$ (rotation  along the $x$-axis  for $\pi$).
			\item[(iii)] ${\hat {\cal T}}$: time reversion.
		\end{itemize}

		In space \(\mathscr{H}\), the total single-molecule Hamiltonian is \(\hat{H}_0(E_0) := \hat{H}_0^{(L)}(E_0) + \hat{H}_0^{(R)}(E_0)\), where $\hat{H}_0^{(L,R)}(E_0)$ were introduced earlier. 
 It can be directly proven that~\cite{SM}, both  \(\hat{H}_0(E_0)\) and 
		  \(\hat{J}_z\) are invariant under the combined transformation \(\hat{\mathcal{P}}\hat{\mathcal{C}}_{2x}\hat{\mathcal{T}}\), regardless of whether $E_0$ is zero or non-zero.
Thus, states \(|\xi, M; R\rangle\) and \(|\xi, M; L\rangle\), which are degenerate common eigenstates of \(\hat{H}_0\) and \(\hat{J}_z\), can be related via 
$
|\xi, M; R\rangle = \hat{\mathcal{P}}\hat{\mathcal{C}}_{2x}\hat{\mathcal{T}}|\xi, M; L\rangle
$~\cite{SM}.
Using this fact and Eqs.~(\ref{apm0}, \ref{bpm0}), we obtain 
$
a_{\pm 1}^{(R)} = -a_{\pm 1}^{(L)^*}$ and $b_{\pm 1}^{(R)} = -b_{\pm 1}^{(L)^*}
$~\cite{SM}.
This result and Eq.~(\ref{tast})  directly  yield that
\(\theta_{\rm f}^{(L)} = \theta_{\rm f}^{(R)}\) only when the parameter \(a_{+}^{(s)^*} b_{-}^{(s)^*} b_{+}^{(s)} a_{-}^{(s)}\) is real.

		Furthermore, as proven in the SM~\cite{SM},
		the  factor $a_{+}^{(s)\ast}
		b_{-}^{(s)\ast}
		b_{+}^{(s)}
		a_{-}^{(s)}$ ($s=L, R$) is real, provided that \(\hat{H}_0\) is invariant under the combined transformation $\hat {\cal C}_{2x}{\hat {\cal T}}$. 
		This condition is satisfied 
		for $E_0=0$, since \(\hat{H}_0(E_0=0)\) is invariant under both $\hat {\cal C}_{2x}$ and ${\hat {\cal T}}$. Thus, 
		$\theta_{\rm f}^{(R)}(E_0=0)= \theta_{\rm f}^{(L)}(E_0=0)$. When $E_0\neq 0$, \(\hat{H}_0(E_0)\) is still invariant under ${\hat {\cal T}}$, but not invariant under $\hat {\cal C}_{2x}$, and thus not invariant under $\hat {\cal C}_{2x}{\hat {\cal T}}$. As a result, 
		the factor   $a_{+}^{(s)\ast}
		b_{-}^{(s)\ast}
		b_{+}^{(s)}
		a_{-}^{(s)}$ is generally not real, and thus we generally have $\theta_{\rm f}^{(R)}\neq \theta_{\rm f}^{(L)}$.
		
Here we also point out that, the aforementioned results $
a_{\pm 1}^{(R)} = -a_{\pm 1}^{(L)^*}$ and $b_{\pm 1}^{(R)} = -b_{\pm 1}^{(L)^*}
$ and Eq.~(\ref{tast}) also yield $\theta_{\rm f}^{(R)}=- \theta_{\rm f}^{(L)}$ (mod $2\pi$). Thus, the degree $D$ of enantiospecificity of the TPSR takes its maximum value (i.e., $D=1$)
only when $|\theta_{\rm f}^{(L,R)}|=\pi/2$, and becomes zero only when 
		$\theta_{\rm f}^{(L,R)}=0$ or $\pi$.

{\it Enantiospecific transitions.}---With the help of the enantiospecific TPSR, one can realize enantiospecific transitions, i.e., the transition from $\alpha$ to $\gamma$,  which occurs exclusively for the enantiomer with a specific handedness. For instance, when the angle $\theta$ of polarization directions of beam 2 is tuned to  $\theta_{\rm f}^{(L)}$, the left-handed molecules can only be transferred from $\alpha$ to $\beta$ by beam 1, and cannot be further transferred from $\beta$ to $\gamma$ by beam 2. In contrast, the right-handed molecules can be transferred from level $\alpha$ to $\beta$ and subsequently to $\gamma$ by the two beams [Fig.~\ref{fig0}(c) and Fig.~\ref{fig4}(b), left]. Therefore, if a molecule is detected at the $\gamma$ level, it is guaranteed to be right-handed. Similarly, another enantiospecific transition can be realized when $\theta$ is tuned to $\theta_{\rm f}^{(R)}$  [Fig.~\ref{fig0}(c) and Fig.~\ref{fig4}(b), right].
Note that these results are independent of the detunings, phases, intensities and operational times of the two beams. As a result, the resonance condition and precise control or locking of these parameters are not
required for the realization of enantiospecific transitions.

\begin{figure}[tbp]
			\includegraphics[width=0.85\columnwidth]{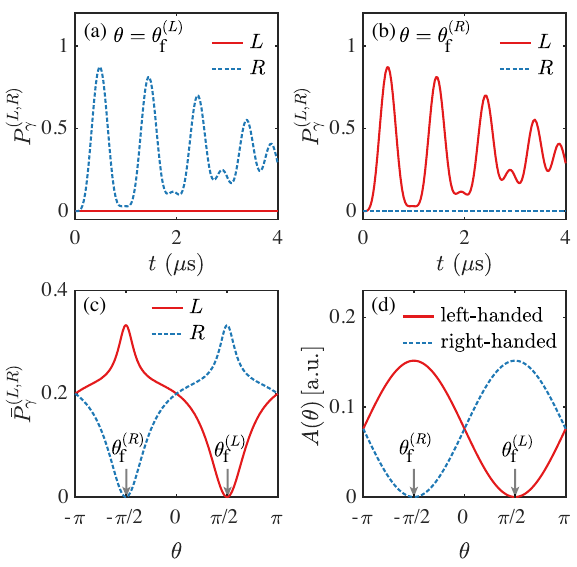}
			\caption{ {\bf (a, b):} Time evolution of the finial-state probability $P^{(L,R)}_\gamma(t)$ of the  cascade transition (\ref{tp}) for $\theta=\theta_{\rm f}^{(L)}$ (a) and $\theta=\theta_{\rm f}^{(R)}$ (b). {\bf (c):} The time-averaged finial-state probability  $\bar P^{(L,R)}_\gamma$ as a function of $\theta$. 
				In (a-c) we consider  the cases with  $\theta_{\rm f}^{(L)}=\pi/2$, $\theta_{\rm f}^{(R)}=-\pi/2$, $\epsilon_{\beta, \pm1}-\epsilon_{\alpha,0}-\omega_{1}={(2\pi)0.1}\,$MHz, $\epsilon_{\gamma,0}-\epsilon_{\beta,\pm 1}-\omega_2={(2\pi)0.4}\,$MHz, and $\Omega_{1}=\Omega_{2}=(2\pi)1\,$MHz, with $\omega_{1(2)}$ and $\Omega_{1(2)}$ being the angular frequency of beam 1(2) and the Rabi frequency of the transition induced by beam 1 (2), respectively.  {\bf (d):} Absorption $A(\theta)$ of beam 2, of left- and right-handed enantiomers. Here we show the results
				 for the system with parameters (expect $\Omega_{2}$) being  same as (c), and the spontaneous emission rates from  $|\gamma,0,s\rangle$ to the lower states $|\alpha,0,s\rangle$ and $|\beta,\pm 1,s\rangle$ all being $\kappa=(2\pi)0.1\,$MHz. The details of  the calculations for (a-d)  are give in the SM~\cite{SM}.}
			\label{result2}
		\end{figure}

		We illustrate the enantiospecific transition by solving the time-dependent Schr\"odinger equation~\cite{SM} of the rovibrational state $|\Psi^{(s)}(t)\rangle$ of molecule $s$ ($s=L,R$)
		for the transition (\ref{tp}), with the
		forbidden polarization angles being $\theta_{\rm f}^{(L/R)}=\pm\pi/2$, and the
		initial state being $|\Psi^{(s)}(0)\rangle=|\alpha, 0; s \rangle$.
		 We show the time evolution of the finial-state probability 
		$P^{(s)}_\gamma(t):=|\langle \gamma,0;s|\Psi(t)\rangle|^2$ ($s=L,R$)
		for the  cases with $\theta=\theta^{(L)}_{\rm f}$ and $\theta=\theta^{(R)}_{\rm f}$
		in Figs.~\ref{result2}(a) and~\ref{result2}(b), respectively.
		As shown in Fig.~\ref{result2}(a),  
		when $\theta=\theta^{(L)}_{\rm f}$, the probability  $P^{(L)}_\gamma(t)$ is always zero, while    
		$P^{(R)}_\gamma(t)$
		still oscillates with time. 
		In another word, the  two-photon transition for the left-handed molecule is totally disabled, while the one for the right-handed molecule is still happening.
		When $\theta=\theta^{(R)}_{\rm f}$ there is a similar result, as shown in Fig.~\ref{result2}(b).
		
		When $\theta$	is not tuned to $\theta_{\rm f}^{(L)}$ or $\theta_{\rm f}^{(R)}$, the $\alpha\leftrightarrow\gamma$ transition can occur for both enantiomers, and thus is not enantiospecific. However, since the  matrix elements $\langle \gamma,\!0 ; s|\hat {\bm d}\cdot{\mathbf{e}}_2| \psi; s\rangle$
		for $s=L$ and $s=R$ are unequal, there are still quantitative difference between
		the finial-state probabilities $\bar P^{(L,R)}_\gamma(t)$ of enantiomers with opposite handedness. We illustrate this in Fig.~\ref{result2}(c), where
		 the time-averaged finial-state probability $\bar P^{(L,R)}_\gamma:=\lim_{T\rightarrow \infty}\frac 1T \int_0^TP^{(L,R)}_\gamma(t)dt$  is plotted as functions of $\theta$, for the above example. 		
				
		If the state $|\gamma,0,s\rangle$ is in a high electronic energy level, one can  verify the 
		 above effects via measuring the absorption  $A$ of  beam 2 by the molecule, in the presence of 
 beam 1 and the static E-field.
When $\theta=\theta_{\rm f}^{(L)} (\theta_{\rm f}^{(R)}$), we have
$A=0$ for the left- (right-) handed enantiomer,  reflecting the enantiospecific transition. 
For other  $\theta$, the absorption $A$ of the two enantiomers are both non-zero, but unequal with each other.
We illustrate these results via Fig.~\ref{result2}(d), where $A$  is plotted as a function of $\theta$ for an example~\cite{SM}.



		{\it  Applications.---}
		The enantiospecific  transitions  based on enantiospecific TPSR can be used for enantiospecific state transfer, enantiodetection, and enantioseparation. 
		For instance,  in a mixture of enantiomers, when the angle $\theta$ is tuned to $\theta_{\rm f}^{(L)}$ ($\theta_{\rm f}^{(R)}$), the absorption of beam 2 by the molecules becomes proportional to the density of the right- (left-) handed enantiomers. Thus, one can derive the enantiomeric excess by detecting these absorptions. 
		Additionally, to realize enantioseparation, one can exclusively  transfer the molecules with a certain handedness to the $\gamma$ level, and then ionize or dissociate  the molecules in the $\gamma$ level via a laser. By recombining these ionized or dissociated products, one can obtain molecules with the certain handedness.

	As mentioned earlier, the resonance condition and phase locking of the two beams, as well as precise control over their intensities and operation times, are not required to realize enantiospecific transitions. These factors significantly simplify the procedure and enhance the feasibility of the aforementioned applications. In particular, since phase-locking is not necessary, the frequencies of the two beams are unrestricted. As a result, the energy levels \(\alpha\), \(\beta\), and \(\gamma\), can be chosen arbitrarily, either within the same electronic or vibrational state or not. Thus, one can select the final state \(\gamma\) to be sufficiently high such that thermal occupation in this state can be safely neglected, even at room temperature.

The enantiospecific transitions occur only when \(\theta\) is tuned to \(\theta_{\rm f}^{(L)}\) or \(\theta_{\rm f}^{(R)}\). However, as shown in Figs.~\ref{result2}(c, d), when \(\theta\) takes other values, the \(\alpha \leftrightarrow \gamma\) transition for enantiomers with different handedness remains quantitatively different. These distinct behaviors can also be utilized for applications such as enantiodetection and enantioseparation. In such cases, controlling beam detuning, intensities or operation times may be necessary, while phase locking is generally still not required. Thus, the advantage of not needing phase locking remains intact.
		

		\begin{acknowledgments}
			This work is supported by the Innovation Program for Quantum Science and Technology (Grant No.~2023ZD0300700),  the National Key Research and Development Program of China (Grant No.~2022YFA1405300), the National Natural Science Foundation of China (Grants No.~12274107 and No.~12405011), and the Natural Science Foundation of Hainan Province (Grants No.~125QN210 and No.~125RC631).
		\end{acknowledgments}

	\section{End Matter}
	
	

\begin{figure}[h]
			\includegraphics[width=0.9\columnwidth]{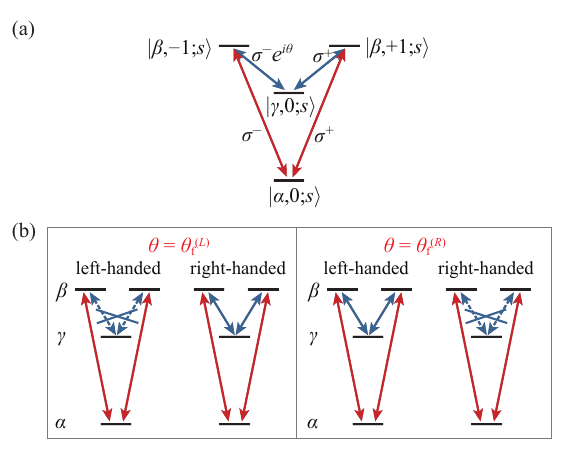}
			\caption{Schematics for $\epsilon_{\alpha, 0}<\epsilon_{\gamma, 0}<\epsilon_{\beta,\pm 1}$.
			{\bf (a):} The rovibrational levels and the circularly-polarized components of beams 1 (red) and 2 (blue) involved in the cascade transition of Eq.~(\ref{tp}). {\bf (b):} Enantiospecific  transitions.
			}
			\label{fig4}
		\end{figure}

{\it Schematics for $\epsilon_{\alpha, 0}<\epsilon_{\gamma, 0}<\epsilon_{\beta,\pm 1}$.} 
In Figs.~\ref{fig0} (b, c) we present the schematics for the two-photon cascade transition of Eq.~(\ref{tp}) and the TPSR for the cases with $\epsilon_{\alpha, 0}<\epsilon_{\beta,\pm 1}<\epsilon_{\gamma, 0}$. Here we give the schematics for another kind of cases, i.e., the ones with $\epsilon_{\alpha, 0}<\epsilon_{\gamma, 0}<\epsilon_{\beta,\pm 1}$, in Fig.~\ref{fig4}.

		\twocolumngrid

	%


		\onecolumngrid
		
		\setcounter{equation}{0}
		\setcounter{figure}{0}
		\setcounter{table}{0}
		\setcounter{page}{1}
		\setcounter{secnumdepth}{3}
		\makeatletter
		\renewcommand{\theequation}{S\arabic{equation}}
		\renewcommand{\thefigure}{S\arabic{figure}}
		\renewcommand{\thesection}{S\arabic{section}}
		\renewcommand{\bibnumfmt}[1]{[S#1]}
		\renewcommand{\citenumfont}[1]{S#1}
		\providecommand{\tabularnewline}{\\}
		
		\begin{center}
			{\large \bf Supplemental Material for \textquotedblleft Enantiospecific Two-Photon Electric-Dipole Selection Rule of Chiral Molecules\textquotedblright}
		\end{center}
		
		\begin{center}
			Fen Zou,$^{1}$ Yong Li,$^{1,*}$ and Peng Zhang$^{2,3,\dagger}$
		\end{center}

		\begin{minipage}[]{18cm}
			\small{\it
				\centering $^{1}$Center for Theoretical Physics \& School of Physics and Optoelectronic Engineering, Hainan University, Haikou 570228, China \\
				\centering $^{2}$School of Physics, Renmin University of China, Beijing, 100872, China\\
				\centering $^{3}$Key Laboratory of Quantum State Construction and Manipulation (Ministry of Education), \\
				Renmin University of China, Beijing, 100872, China\\}
		\end{minipage}
		\vspace{8mm}

		\section{Calculation for Results of Fig.~2} 
		\label{a1}
		
		In this section we show our approach for the calculation of 
		the forbidden polarization angle $\theta_{\rm f}^{(L,R)}(E_0)$ of the left- and right-handed 1,2-propanediol molecules, i.e., the results of Fig.~\ref{result} in the main text.
		
		We first derive the results for the left-handed molecule, and then derive the results for the right-handed one.		
		
		\subsection{The Left-Handed Molecule ($s=L$)}
		
		We first derive the eigen-states $|\xi,\!M;\!L\rangle$ ($\xi=1,2,...; M=0,\pm 1$) and the corresponding eigen-energies $\epsilon_{\xi, M}$
		of the Hamiltonian $\hat H^{(L)}_0(E_0)$ of molecule $L$ in the static E-field ${\bm E}=E_0{\mathbf e}_z$.
		This Hamiltonian can be expressed as
		\begin{eqnarray}
			\hat H^{(L)}_0(E_0)=\hat H^{(L)}_{\rm F}-E_0{\hat {\bm d}}\cdot{\mathbf{e}}_z,\label{h0sa}
		\end{eqnarray}
		where $\hat H^{(L)}_{\rm F}$ is the Hamiltonian  in the absence of the static E-field, and 		$\hat  {\bm d}$ is the molecular electric-dipole operator.

		As mentioned in the main text, in this work we  assume $E_0$ is weak enough so that we only require to consider the $E_0$-induced
		coupling between
		the eigen-states of $\hat H^{(L)}_{\rm F}$ (i.e., the rotational states) in the same electronic and vibrational level. On the other hand, as shown in the caption of Fig.~\ref{result}, we focus on several lowest eigen-states of $\hat H^{(L)}_0(E_0)$ (i.e., the states with quantum numbers $\alpha,\beta,\gamma<4$).
		Due to these facts, in the calculation we only consider the rotational states in the lowest electronic and vibrational level. 
		
		For the convenience of the following calculation, we denote the three principal axes of inertia of the left-handed molecule as $a$, $b$ and $c$, with corresponding  moments of inertia $I_{a}$, $I_{b}$ and $I_{c}$. Without loss of generality, we assume $I_{a}<I_{b}, I_{c}$, and the axes $a$, $b$ and $c$ form a right-handed frame. We further introduce the basis $\{|J,K,M;L\rangle\}$ of the rotational states of the lowest electronic and vibrational level. Here 
		$J=0,1,2,...$ is the quantum number of $\hat {\bm J}^2$, with $\hat {\bm J}$ being the total angular momentum, as mentioned above, and $M$ and $K$ are quantum numbers of 
		of the angular momentum along the $z$-axis of the lab frame (i.e., $\hat J_z$) and the one along the principal axis $a$ (i.e., $\hat J_a$), respectively.

		To derive the eigen-states and eigen-values of $\hat H^{(L)}_0(E_0)$,
		we express the terms in the right-hand-side of Eq.~(\ref{h0sa})
		as matrices in the basis  $\{|J,K,M;L\rangle\}$. 
		The matrix elements of $\hat H^{(L)}_{\rm F}$ in this basis are given by~\cite{Zare1988SM}
		\begin{eqnarray}
			\langle J,K,M;L|\hat H^{(L)}_{\rm F}|J', K', M';L\rangle&=&\delta_{J,J'}\delta_{M,M'}\bigg[
			f(J,K)\delta_{K',K}+g_{\pm}(J,K)\delta_{K',K\pm 2}
			\bigg];\label{me1}
		\end{eqnarray}
		where $\delta_{\alpha,\beta}$ is the Kronecker delta symbol, and the functions $f(J,K)$
		and $g^{(\pm)}(J,K)$ are defined as
		\begin{eqnarray}
			f(J,K)&=&\frac 12(B+C)\left[J(J+1)-K^2\right]+AK^2;\\
			\nonumber\\
			g_{\pm}(J,K)&=&\frac 14(B-C)\Bigg\{\bigg[J(J+1)-K(K\pm 1)\bigg]
			\bigg[J(J+1)-(K\pm 1)(K\pm 2)\bigg]\Bigg\}^{1/2},
		\end{eqnarray} 
		with
		\begin{eqnarray}
			A=\frac{\hbar^2}{2I_{a}},\ B=\frac{\hbar^2}{2I_{b}},\ C=\frac{\hbar^2}{2I_{c}}.
		\end{eqnarray} 
		For a left-handed 1,2-propanediol molecule we have
		$A=\hbar(2\pi)8572.05\,\textrm{MHz}$, $B=\hbar(2\pi)3640.10\,\textrm{MHz}$, and $C=\hbar(2\pi)2790.96\,\textrm{MHz}$~\cite{Lovas2009SM}.
		Additionally, the matrix element of the term ${\hat {\bm d}}\cdot{\mathbf{e}}_z$ in the basis $\{|J,K,M;L\rangle\}$ can be expressed as~\cite{Jacob2012SM,Ye2018SM}
		\begin{eqnarray} 
			&&\langle J,K,M;L|\big({\hat {\bm d}}\cdot{\mathbf{e}}_z\big)|J', K', M';L\rangle\nonumber\\
			\nonumber\\
			&=&
			\delta_{M,M'}
			\sqrt{(2J+1)(2J'+1)}
			\left(
			\begin{array}{ccc}
				J&1&J'\\
				M&0&-M'
			\end{array}
			\right)
			\Bigg\{
			\sum_{\sigma'=0,\pm 1}
			\mu^{(m)}_{\sigma'}
			(-1)^{M'+1-K'+\sigma'}
			\left(
			\begin{array}{ccc}
				J&1&J'\\
				K&-\sigma'&-K'
			\end{array}
			\right)
			\Bigg\},
			\label{me2}
		\end{eqnarray} 
		with $\mu^{(m)}_{0,\pm 1}$ being given by
		\begin{eqnarray}
			\mu^{(m)}_{0}=-d_a;\hspace{0.5cm}
			\mu^{(m)}_{\pm}=\pm(d_b\pm id_c)/\sqrt{2},
		\end{eqnarray}
		where $(d_a,d_b,d_c)$ are components of the electric dipole moment of the molecule $L$ in the principle-axis frame ($abc$-frame), which are independent of $(K,K',M,M')$. For a left-handed 1,2-propanediol molecule we have $d_{a}=-1.201\,\textrm{Debye}$, $d_{b}=-1.916\,\textrm{Debye}$, and $d_{c}=-0.365\,\textrm{Debye}$~\cite{Lovas2009SM}.
		
		Using Eqs.~(\ref{me1}, \ref{me2}) and Eq.~(\ref{h0sa}), we can 
		express $\hat H^{(L)}_0(E_0)$ as a matrix in the basis $\{|J,K,M;L\rangle\}$. We
		derive the eigen-states $\{|\xi,\!M;\!L\rangle\}$ and eigen-energies $\{\epsilon_{\xi, M}\}$ by directly diagonalizing this matrix. Specifically, the eigen-states $\{|\xi,\!M;\!L\rangle\}$ can be expressed as
		\begin{eqnarray}
			|\xi,\!M;\!L\rangle=\sum_{J=|M|}^{+\infty}\sum_{K=-J}^{+J}
			S_{\xi,J,K;M}^{(L)}|J,K,M;L\rangle,\label{st}
		\end{eqnarray}
		while the coefficients $\{S_{\xi,J,K;M}^{(L)}\}$ are given by numerical diagonalization of the matrix.
		
		Substituting the states $|\xi,\!M;\!L\rangle$ of Eq.~(\ref{st}), which are derived via the above approach, into Eqs.~(\ref{apm0}, \ref{bpm0}), and using the relations~\cite{Jacob2012SM,Ye2018SM}
		\begin{eqnarray} 
			&&\langle J,K,M;L|\big({\hat {\bm d}}\cdot{\mathbf{e}}_\pm\big)|J', K', M';L\rangle\nonumber\\
			\nonumber\\
			&=&
			\delta_{M,M'\pm 1}
			\sqrt{(2J+1)(2J'+1)}
			\left(
			\begin{array}{ccc}
				J&1&J'\\
				M&\mp 1&-M'
			\end{array}
			\right)
			\Bigg\{
			\sum_{\sigma'=0,\pm 1}
			\mu^{(m)}_{\sigma'}
			(-1)^{M'-K'+\sigma'}
			\left(
			\begin{array}{ccc}
				J&1&J'\\
				K&-\sigma'&-K'
			\end{array}
			\right)
			\Bigg\},
			\label{me3}
		\end{eqnarray} 
		we directly obtain the coefficients $
		a_{\pm }^{(L)}$ and $b_{\pm}^{(L)}
		$.
		
		Finally, we submit the coefficients $
		a_{\pm }^{(L)}$ and $b_{\pm}^{(L)}$ derived above into Eq.~(\ref{tast}), and obtain the forbidden polarization angle $\theta_{\rm f}^{(L)}(E_0)$.
		
		\subsection{The Right-Handed Molecule ($s=R$)}
		
		Now we consider the right-handed 1,2-propanediol molecule. We derive the forbidden polarization angle $\theta_{\rm f}^{(R)}(E_0)$ of this molecule with the same method as above. 
		
		To this end, we introduce the rotational states $\{|J,K,M;R\rangle\}$ of the right-handed molecule in the lowest electric and vibrational level, which are defined as
		\begin{eqnarray}
			|J,K,M;R\rangle=\hat {\cal P}|J,K,M;L\rangle,
		\end{eqnarray}
		where $\hat {\cal P}$ is the spatial inversion operator for  the coordinates of all nuclei and electrons in the molecule. 
		
		Furthermore, 
		similar as in the above subsection, we
		express the  Hamiltonian 
		$\hat H^{(R)}_0(E_0)=\hat H^{(R)}_{\rm F}-E_0{\hat {\bm d}}\cdot{\mathbf{e}}_z$ and
		the operators ${\hat {\bm d}}\cdot{\mathbf{e}}_\pm$
		as matrices 
		in the basis 
		$\{|J,K,M;R\rangle\}$.
		Note that 
		since $\hat {\cal P}\hat H^{(R)}_{\rm F}\hat {\cal P}=\hat H^{(L)}_{\rm F}$ and 
		$\hat {\cal P}({\hat {\bm d}}\cdot{\mathbf{e}}_{{z},\pm})\hat {\cal P}=-{\hat {\bm d}}\cdot{\mathbf{e}}_{{z},\pm}$, we have 
		\begin{eqnarray}
			\langle J,K,M;R|\hat H^{(R)}_{\rm F}|J', K', M';R\rangle&=&+\langle J,K,M;L|\hat H^{(L)}_{\rm F}|J', K', M';L\rangle,\label{rel1}\\
			\nonumber\\
			\langle J,K,M;R|\big({\hat {\bm d}}\cdot{\mathbf{e}}_{{z},\pm}\big)|J', K', M';R\rangle
			&=&-\langle J,K,M;L|\big({\hat {\bm d}}\cdot{\mathbf{e}}_{{z},\pm}\big)|J', K', M';L\rangle.\label{rel2}
		\end{eqnarray}
		We calculate  the matrix elements 
		for the right-handed molecule by substituting the results (\ref{me1}, \ref{me2}, \ref{me3})
		into  Eqs.~(\ref{rel1}, \ref{rel2}).
		
		After  expressing
		$\hat H^{(R)}_0(E_0)$ and
		${\hat {\bm d}}\cdot{\mathbf{e}}_\pm$
		as matrices 
		in the basis 
		$\{|J,K,M;R\rangle\}$, we calculate the eigen-states $\{|\xi, M,R\rangle\}$ and the eigen-energies $\{\epsilon_{\xi, M}\}$ of $\hat H^{(R)}_0(E_0)$, and then derive the parameters $a_{\pm}^{(R)}$, $b_{\pm}^{(R)}$, and the forbidden polarization angle $\theta_{\rm f}^{(R)}{(E_{0})}$, with the exact same method as in the above subsection.
		
		\section{$\theta_{\rm f}^{(L,R)}(E_0)$ and $D(E_0)$ for more cases}
		
\begin{figure}[h]
			\includegraphics[width=0.5\columnwidth]{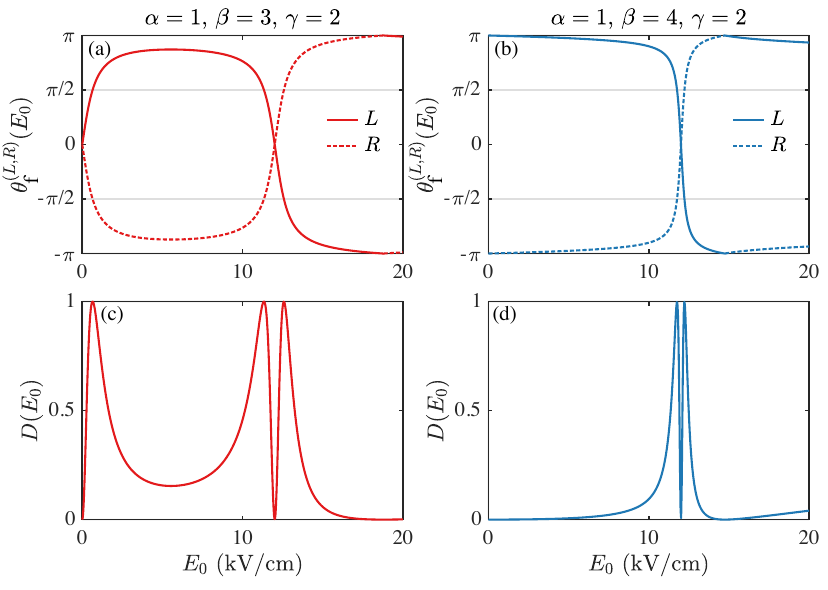}
			\caption{{\bf (a, b):} The forbidden polarization angle $\theta_{\rm f}^{(L,R)}(E_0)$. {\bf (c, d):} The  degree  $D(E_{0})$  of enantiospecificity
				of the TPSR. Here we show the results for  transitions of Eq.~(\ref{tp}) for 1,2-propanediol molecules, with relevant quantum numbers being $(\alpha=1;\beta=3;\gamma=2)$ (a, c), and $(\alpha=1;\beta=4;\gamma=2)$ (b, d).}
			\label{FigS1}
		\end{figure}

		In Fig.~\ref{result} of our main text we show the forbidden polarization angles $\theta_{\rm f}^{(L,R)}$ and the degree of enantiospecificity $D$ as functions of the static E-field strength $E_0$,
for 1,2-propanediol molecules with relevant quantum numbers $(\alpha=1;\beta=1;\gamma=4)$ (a, c), and $(\alpha=1;\beta=3;\gamma=4)$. Here we illustrate $\theta_{\rm f}^{(L,R)}(E_0)$ and $D(E_0)$ for other two cases with $(\alpha=1;\beta=3;\gamma=2)$ and $(\alpha=1;\beta=4;\gamma=2)$ in Fig.~\ref{FigS1}.
		
				\section{ Analysis on the origin of the enantiospecific  TPSR}
		
		In this section we show some details of the analysis on the origin of the enantiospecific  TPSR.
		
		\subsection{Hamiltonian and Hilbert Space}
		
		As in the maintext, we denote $x,y,z$ as the spatial coordinates of the particles (atomic nucleus and electrons) of each molecule, in the center-of-mass  (CoM) frame of this molecule, e.g., $x:=\{x_1,x_2,...\}$, where $x_{i}$ ($i=1,2,...$) is the $x$-coordinate the particle $i$. Note that the origin of the CoM frame is located at the CoM position of the molecule, and the axes of this frame are parallel to those of the lab frame.
		Each molecular state
		$|\xi,\!M;\!s\rangle$
		is just a state of the relative motion of these  particles, and corresponds to a wave function $\langle x,y,z|\xi,\! M;\!s\rangle$.
		We further  denote $\mathscr H_{s}$ ($s=L,R$) as the Hilbert space spanned by  $\{|\xi,\! M;\!s\rangle\vert_{E_0=0}\}$, i.e., the space of the states with handedness $s$. 
		
		
		

		In the total Hilbert space \(\mathscr{H}=\mathscr H_{L}\oplus \mathscr H_{R}\),
		  the Hamiltonian of the molecular internal state is
		\begin{eqnarray}
			\hat H_0(E_0)=\hat H_{\rm F}-E_0{\hat {\bm d}}\cdot{\mathbf{e}}_z,\label{hexa}
		\end{eqnarray}
		where ${\hat {\bm d}}$ is the molecular electric dipole operator, and $\hat H_{\rm F}$ is the {Hamiltonian} in the absence of the static electric field, which includes the relative kinetic energy and interaction of the particles in the molecule.
	Additionally, $\hat H_{\rm F}$ is close in the subspaces $\mathscr H_{L}$ and $\mathscr H_{R}$, and can be expressed as $\hat H_{\rm F}=\hat H_{\rm F}^{(L)}+\hat H_{\rm F}^{(R)}$, where $\hat H_{\rm F}^{(s)}$ ($s=L, R$) operates on the states in the subspace $\mathscr H_{s}$, and is studied in detail in Sec.~\ref{a1}.

		Precisely speaking, the term $-E_0{\hat {\bm d}}\cdot{\mathbf{e}}_z$ in Eq.~(\ref{hexa}) can induce not only the coupling between the states within the specific space  $\mathscr{H}_{L(R)}$, but also the coupling between  $\mathscr{H}_{L}$ and  $\mathscr{H}_{R}$. However, as mentioned in our main text,
we assume that $E_0$ is so weak that we can only consider the 
induced  coupling  between eigen-states of $\hat H_{\rm F}^{(L)}$ ($\hat H_{\rm F}^{(R)}$)
within the same electronic and vibrational level. Clearly, under this approximation, 		
		 the  coupling between  $\mathscr{H}_{L}$ and  $\mathscr{H}_{R}$ is totally ignored, and we can express \(\hat{H}_0(E_0)\) as \(\hat{H}_0(E_0) = \hat{H}_0^{(L)}(E_0) + \hat{H}_0^{(R)}(E_0)\), as in the main text. Here $\hat{H}_0^{(s)}(E_0)$ ($s=L, R$) operates  on the states in the subspace $\mathscr H_{s}$.
		 Accordingly, the eigen-state $|\xi,\!M;\!s\rangle$ of  $\hat H_0^{(s)}(E_0)$, where are in the subspace $\mathscr{H}_{s}$,
are also eigen-state of  ${\hat H}_0(E_0)$.

		\subsection{Properties of the Transformations}
		For convenience of the following discussions, here we show some properties of the transformations ${\hat {\cal P}}$, ${\hat {\cal C}_{2x}}$ and ${\hat {\cal T}}$ introduced in the main text.

		(1) About the free Hamiltonian $\hat H_{\rm F}$:
		\begin{eqnarray}
			{\hat {\cal C}_{2x}}\hat H_{\rm F}{\hat {\cal C}_{2x}}={\hat {\cal P}}\hat H_{\rm F}{\hat {\cal P}}={\hat {\cal T}}\hat H_{\rm F}{\hat {\cal T}}=\hat H_{\rm F}.\label{t1}
		\end{eqnarray}
		
		(2) About the dipole operator:
		\begin{eqnarray}
			{\hat {\cal P}}\big(\hat {\bm d}\cdot{\bf e}_{z,\pm}\big){\hat {\cal P}}&=&-\hat {\bm d}\cdot{\bf e}_{z,\pm};\label{t2}\\[0.2cm]
			{\hat {\cal C}_{2x}}\big(\hat {\bm d}\cdot{\bf e}_z\big){\hat {\cal C}_{2x}}&=&
			-\hat{\bm d}\cdot{\bf e}_{z};\\[0.2cm]
			{\hat {\cal T}}\big(\hat {\bm d}\cdot{\bf e}_z\big){\hat {\cal T}}&=&
			\hat{\bm d}\cdot{\bf e}_{z};\label{t3}
			\\[0.2cm]
			{\hat {\cal C}_{2x}}\big(\hat {\bm d}\cdot{\bf e}_\pm\big){\hat {\cal C}_{2x}}&=&
			{\hat {\cal T}}\big(\hat {\bm d}\cdot{\bf e}_{\pm}\big){\hat {\cal T}}
			=
			-\hat {\bm d}\cdot{\bf e}_\mp.\label{t4}
		\end{eqnarray}
		
		(3) We define
		$ |{\cal T}[\xi,M;s]\rangle$ as the result of applying the time-reversal operator $\hat{\cal T}$ to state $|\xi,M;s\rangle$, i.e.,
		\begin{eqnarray}
			|{\cal T}[\xi,M;s]\rangle:= \hat{\cal T}|\xi,M;s\rangle,
			\hspace{1cm}
			(\xi=1,2,...; M=0,\pm 1; s=L,R).\label{t5}
		\end{eqnarray}
Thus,  
		\begin{eqnarray}
			\langle{\bm r}|{\cal T}[\xi,M;s]\rangle=\langle{\bm r}|\xi,M;s\rangle^\ast,
			\hspace{1cm}
			(\xi=1,2,...; M=0,\pm 1; s=L,R).\label{t6}
		\end{eqnarray}

		\subsection{Proof of the Relations $
			a_{\pm }^{(R)}=-a_{\pm }^{(L)^\ast}$ and $b_{\pm }^{(R)}=-b_{\pm }^{(L)^\ast}
			$}
		
		In this subsection we prove 
		the relations $
		a_{\pm }^{(R)}=-a_{\pm }^{(L)^\ast}$ and $b_{\pm }^{(R)}=-b_{\pm }^{(L)^\ast}
		$, which are used to derive
		Eq.~(\ref{tast}). As mentioned in the main text, the states
		$|\xi,\! M;\!R\rangle$ and  $|\xi,\! M;\!L\rangle$ are related via 
		$
		|\xi,\! M;\!R\rangle=\hat {\cal P}{\hat {\cal C}}_{2x}\hat {\cal T}|\xi,\! M;\!L\rangle.
		$
		Substituting this result into Eq.~(\ref{apm0}), we obtain
		\begin{eqnarray}
			a_{+}^{(R)}
			&=&\langle \beta,+1;R|\hat {\bm d}\cdot{\mathbf{e}}_+| \alpha,0; R\rangle=\int d{\bm r} 
			\langle{\cal T}[\beta,\! +1;\!L]|{\bm r}
			\rangle\bigg[{{\hat {\cal C}}_{2x}\hat {\cal P}\big(\hat {\bm d}\cdot{\mathbf{e}}_+\big)\hat {\cal P}{\hat {\cal C}}_{2x}}\bigg]
			\langle{\bm r}|{\cal T}[\alpha,\! 0;\!L]\rangle,
			\label{pro}
		\end{eqnarray} 
		On the other hand, Eqs.~(\ref{t2}, \ref{t4}) yield
		\begin{eqnarray}
			{{\hat {\cal C}}_{2x}\hat {\cal P}\big(\hat {\bm d}\cdot{\mathbf{e}}_+\big)\hat {\cal P}{\hat {\cal C}}_{2x}}=\hat {\bm d}\cdot{\mathbf{e}}_-.
		\end{eqnarray}
		Substituting this result and Eq.~(\ref{t6}) and $\mathbf{e}_-=-\mathbf{e}_+^\ast$
		into Eq.~(\ref{pro}), and then obtain
		\begin{eqnarray}
			a_{+}^{(R)}
			&=&\left[-\int d{\bm r} \langle\beta,\! +1;\!L|{\bm r}
			\rangle\big(\hat {\bm d}\cdot{\mathbf{e}}_+\big)
			\langle{\bm r}|\alpha,\! 0;\!L\rangle\right]^\ast=-a_{+}^{(L)\ast}.
		\end{eqnarray} 
		Thus, we have proven $a_{+}^{(R)}=-a_{+}^{(L)\ast}$. The relations $a_{-}^{(R)}=-a_{-}^{(L)\ast}$ and $b_{\pm}^{(R)}=-b_{\pm}^{(L)\ast}$ can be proven with the same approach.
		
		\subsection{Relation Between the Factor $a_{+}^{(s)\ast}
			b_{-}^{(s)\ast}
			b_{+}^{(s)}
			a_{-}^{(s)}$ and the Transformation $\hat {\cal C}_{2x}{\hat {\cal T}}$ }
		
		In this {subsection} we prove that  the factor $a_{+}^{(s)\ast}
		b_{-}^{(s)\ast}
		b_{+}^{(s)}
		a_{-}^{(s)}$ is real when the Hamiltonian $\hat H_0^{(s)}(E_0)$ ($s=L,R$) is invariable under the combined transformation $\hat {\cal C}_{2x}{\hat {\cal T}}$, i.e., when $[\hat H_0^{(s)}(E_0), \hat {\cal C}_{2x}{\hat {\cal T}}]=0$.
		
		We notice that both the operators $\hat H_0^{(s)}(E_0)$ and $\hat {\cal C}_{2x}{\hat {\cal T}}$ are close in the subspace with a fixed quantum number $M$ of the total angular momentum $\hat J_z$, and in this subspace $\hat H_0^{(s)}(E_0)$ does not degenerate. Due to these facts,
		when $[\hat H_0^{(s)}(E_0), \hat {\cal C}_{2x}{\hat {\cal T}}]=0$, by choosing a proper global phase, we can always make the 
		the states $|\xi,\!M;\!s\rangle$ ($\xi=1,2,...; M=0,\pm 1; s=L,R$) to be an eigen-state of $\hat {\cal C}_{2x}{\hat {\cal T}}$, with eigen-value being $+1$. Namely, we have 
		\begin{eqnarray}
			\hat {\cal C}_{2x}{\hat {\cal T}}|\xi,\!M;\!s\rangle=|\xi,\!M;\!s\rangle. (s=L, R)\label{ctps}
		\end{eqnarray}
		Using this result and Eqs.~(\ref{t4}, \ref{t6}), we can further obtain
		\begin{eqnarray}
			a_{+}^{(s)}
			&=&\langle \beta,+1;s|\hat {\bm d}\cdot{\mathbf{e}}_+| \alpha,0; s\rangle\nonumber\\
			&=&\int d{\bm r} \langle{\cal T}[\beta,\! +1;\!s]|{\bm r}
			\rangle\bigg[{\hat {\cal C}}_{2x}\big(\hat {\bm d}\cdot{\mathbf{e}}_+\big){\hat {\cal C}}_{2x}\bigg]
			\langle{\bm r}|{\cal T}[\alpha,\! 0;\!s]\rangle\nonumber\\
			&=&\left[\int d{\bm r} \langle\beta,\! +1;\!s|{\bm r}
			\rangle\big(\hat {\bm d}\cdot{\mathbf{e}}_+\big)
			\langle{\bm r}|\alpha,\! 0;\!s\rangle\right]^\ast\nonumber\\
			&=&a_{+}^{(s)\ast}.\label{aps}
		\end{eqnarray} 
		Eq.~(\ref{aps}) yields that $a_{+}^{(s)}$ is {\it real}. With the same approach we can also prove that  $a_{-}^{(s)}$ and $b_{\pm}^{(s)}$ are all real.  Therefore, the factor $a_{+}^{(s)\ast}
		b_{-}^{(s)\ast}
		b_{+}^{(s)}
		a_{-}^{(s)}$ is real. 
		
		On the other hand, the value of  $a_{+}^{(s)\ast}
		b_{-}^{(s)\ast}
		b_{+}^{(s)}
		a_{-}^{(s)}$ is independent of the choice of the global phases of the states $|\xi,\!M;\!s\rangle$ ($\xi=1,2,...; M=0,\pm 1; s=L,R$). Therefore, although the above derivation is based on the global phases satisfying Eq.~(\ref{ctps}), it actually leads to the conclusion that $a_{+}^{(s)\ast}
		b_{-}^{(s)\ast}
		b_{+}^{(s)}
		a_{-}^{(s)}$ is always real under the condition $[\hat H_0^{(s)}(E_0), \hat {\cal C}_{2x}{\hat {\cal T}}]=0$.

		\section{Calculation for results of Fig.~3}
		
		Now we show our approach for the calculation of the cascade-transition dynamics, i.e., the results of Fig.~\ref{result2}. 
		
		\subsection{The Finial-State Probability $P_{\gamma}^{(s)}$}
		\label{s3s}
		
		As mentioned in the main text, we consider the case with $\epsilon_{\alpha, 0}<\epsilon_{\beta,\pm1}<\epsilon_{\gamma, 0}$.
		Moreover, the electric field strength ${ \bm {\mathcal  E}}_{1(2)}$ of beams 1 and 2 can be expressed as 
		\begin{eqnarray}
			{ \bm {\mathcal  E}}_{1}={\mathcal  E}_1e^{-i\omega_1t}\mathbf{e}_y+c.c.;\ \ \  { \bm {\mathcal  E}}_{2}={\mathcal  E}_2e^{-i\omega_2t}\mathbf{e}_2+c.c.,
		\end{eqnarray}
		where ${\mathcal  E}_{1,2}$ are corresponding complex amplitudes. 
		When these beams are applied, in the rotating frame
		the  Hamiltonian of the molecule $s$ ($s=L,R$) is given by ($\hbar=1$):
		\begin{eqnarray}
			{\hat H}^{(s)}&=&\Delta_{1}\sum_{M=\pm 1}|\beta,\! M;\!s\rangle\langle\beta,\! M;\!s|+(\Delta_{1}+\Delta_{2})|\gamma,\! 0;\!s\rangle\langle
			\gamma,\! 0;\!s|+\hat H_{\rm I1}+\hat H_{\rm I2},\label{hsapp}
		\end{eqnarray}
		where $\Delta_{1}=\epsilon_{\beta,1}-\epsilon_{\alpha,0}-\omega_1$ and $\Delta_{1}+\Delta_{2}=\epsilon_{\gamma,0}-\epsilon_{\alpha,0}-(\omega_1+\omega_2)$ are the one-photon and two-photon detunings, respectively, with $\omega_{1(2)}$ being the 
		angular frequency of the beam 1(2).
		Additionally, $\hat H_{\rm I1(2)}$ is the interaction between the beam 1 (2) and the molecule, and can be expressed as
		\begin{eqnarray}
			\hat H_{\rm I1}=-\frac{{\cal E}_1}{\sqrt{2}}\left\{a_{+}^{(s)}|\beta,\! +1;\!s\rangle\langle\alpha,\!0;\!s|
			+a_{-}^{(s)}|\beta,\! -1;\!s\rangle\langle\alpha,\!0;\!s|\right\}+h.c.,
		\end{eqnarray}
		and
		\begin{eqnarray}
			\hat H_{\rm I2}&=&-\frac{{\cal E}_2}{\sqrt{2}}\bigg\{b_{-}^{(s)}
			e^{i\theta}
			|
			\gamma,\! 0;\!s
			\rangle\langle\beta,\! +1;\!s|\bigg.+b_{+}^{(s)}|
			\gamma,\! 0;\!s
			\rangle\langle\beta,\! -1;\!s| \bigg\}\!+\!h.c..
		\end{eqnarray}
		Note that $\hat H_{\rm I2}$ depends on the angle $\theta$, due to the fact $\mathbf{e}_2\propto (\mathbf{e}_++e^{i\theta}\mathbf{e}_-)$.
		
		For the convenience of the following calculations, we define the Rabi frequencies:
		\begin{eqnarray}
			\Omega_1=&\sqrt{2}\left\vert
			{\cal E}_1a_+^{(s)}
			\right\vert
			=\sqrt{2}
			\left\vert
			{\cal E}_1a_-^{(s)}
			\right\vert;
			\ \ \ \ \ \ \ \ \ \ \ 
			\Omega_2=\sqrt{2}\left\vert
			{\cal E}_2b_+^{(s)}
			\right\vert
			=\sqrt{2}
			\left\vert
			{\cal E}_2b_-^{(s)}
			\right\vert.
		\end{eqnarray}
		Here we have used the facts $|a^{(L)}_+|=|a^{(L)}_-|=|a^{(R)}_+|=|a^{(R)}_-|$, and $|b^{(L)}_+|=|b^{(L)}_-|=|b^{(R)}_+|=|b^{(R)}_-|$, which are due to the
		the time-reversal symmetry and the relations $
		a_{\pm }^{(R)}=-a_{\pm }^{(L)^\ast}$ and $b_{\pm }^{(R)}=-b_{\pm }^{(L)^\ast}
		$ proven in the main text.
		Note that the Rabi frequencies $\Omega_1$ and $\Omega_2$ are all independent of the molecular handedness. We further define the phases
		\begin{eqnarray}
			\phi_{1\pm }^{(s)}={\rm arg}\left[-{\cal E}_1a_{\pm}^{(s)}\right];\ \ \ \ \ \ \ 
			\phi_{2\pm }^{(s)}={\rm arg}\left[-{\cal E}_2b_{\pm}^{(s)}\right].
		\end{eqnarray}
		Using Eq.~(\ref{tast}) of the main text, we find that these angles are related to the forbidden polarization angle $\theta_{\rm f}^{(s)}$ via
		\begin{eqnarray}
			e^{i\left(\phi^{(s)}_{2+}+\phi^{(s)}_{1-}-\phi^{(s)}_{1+}-\phi^{(s)}_{2-}\right)}=-e^{i\theta_{\rm f}^{(s)}(E_0)}.\label{s19}
		\end{eqnarray}
		
		We further introduce the states
		\begin{eqnarray}
			|\overline{\beta,\pm 1;s}\rangle&=&|\beta,{\pm}1;s\rangle e^{i\phi^{(s)}_{1\pm}};
			\\[0.2cm]
			|\overline{\gamma,0;s}\rangle&=&|\gamma,0;s\rangle 
			e^{i\left(\phi^{(s)}_{1+}{+}\phi^{(s)}_{2-}\right)}.
		\end{eqnarray}
		In the basis $\{|\alpha,0;s\rangle, |\overline{\beta,\pm 1;s}\rangle, |\overline{\gamma,0;s}\rangle\}$, the total Hamiltonian ${\hat H}^{(s)}$ of Eq.~(\ref{hsapp}) can be re-written as
		\begin{eqnarray}
			{\hat H}^{(s)}&=&\Delta_{1}\sum_{M=\pm 1}|\overline{\beta,\! M;\!s}\rangle\langle
			\overline{\beta,\! M;\!s}|+(\Delta_{1}+\Delta_{2})|\overline{\gamma,\! 0;\!s}\rangle\langle
			\overline{\gamma,\! 0;\!s}|\nonumber
			\\[0.4cm]
			&&+\frac{\Omega_1}{2}\bigg[|\overline{\beta,\! +1;\!s}\rangle\langle\alpha,\!0;\!s|
			+|\overline{\beta,\! -1;\!s}\rangle\langle\alpha,\!0;\!s|\bigg]
			+\frac{\Omega_2}{2}\bigg[e^{i\theta}|\overline{\gamma,\! 0;\!s}\rangle\langle\overline{\beta,\!+1;\!s}|
			-e^{i\theta_{\rm f}^{(s)}(E_0)}|\overline{\gamma,\! 0;\!s}\rangle\langle\overline{\beta,\!-1;\!s}|\bigg]+h.c.,\nonumber\\[0.3cm]
			&&\hspace{13cm}(s=L,R).\label{hs2app}
		\end{eqnarray}
		Here we have used  Eq.~(\ref{s19}).
		
				In our calculation, we express the time-dependent state $|\Psi^{(s)}(t)\rangle$ of molecule $s$  ($s=L,R$) as 
		\begin{eqnarray}
			|\Psi^{(s)}(t)\rangle=c^{(s)}_\alpha(t)|\alpha,0;s\rangle+c^{(s)}_{\beta+}(t)|\overline{\beta,+1;s}\rangle+c^{(s)}_{\beta-}(t)|\overline{\beta,-1;s}\rangle+c^{(s)}_\gamma(t)|\overline{\gamma,0;s}\rangle.
		\end{eqnarray}
		Substituting this expression and Eq.~(\ref{hs2app}) into the Scri\"odinger equation
		\begin{eqnarray}
			i\frac{d}{dt}|\Psi^{(s)}(t)\rangle={\hat H}^{(s)}|\Psi^{(s)}(t)\rangle,
		\end{eqnarray}
		we can obtain the equations for the coefficients $c^{(s)}_{\alpha,\beta\pm,\gamma}(t)$ as
		\begin{eqnarray}
			\dot{c}_{\alpha}^{(s)}&=&-i\frac{\Omega_{1}}{2}c_{\beta+}^{(s)}-i\frac{\Omega_{1}}{2}c_{\beta-}^{(s)}, \\
			\dot{c}_{\beta+}^{(s)}&=&-i\frac{\Omega_{1}}{2}c_{\alpha}^{(s)}-i\Delta_{1}c_{\beta+}^{(s)}-i\frac{\Omega_{2}}{2}e^{-i\theta}c_{\gamma}^{(s)},  \\
			\dot{c}_{\beta-}^{(s)}&=&-i\frac{\Omega_{1}}{2}c_{\alpha}^{(s)}-i\Delta_{1}c_{\beta-}^{(s)}+i\frac{\Omega_{2}}{2}e^{-i\theta_{\rm f}^{(s)}}c_{\gamma}^{(s)},  \\
			\dot{c}_{\gamma}^{(s)}&=&-i\frac{\Omega_{2}}{2}e^{i\theta}c_{\beta+}^{(s)}+i\frac{\Omega_{2}}{2}e^{i\theta_{\rm f}^{(s)}}c_{\beta-}^{(s)}-i(\Delta_{1}+\Delta_{2})c_{\gamma}^{(s)}.
		\end{eqnarray}
		Solving this equation with initial condition $c^{(s)}_\alpha(0)=1$ and $c^{(s)}_{\beta\pm,\gamma}(0)=0$ (corresponding to the initial state $|\Psi^{(s)}(0)\rangle=|\alpha,0;s\rangle$), we can obtain the time evolution of the coefficient $c^{(s)}_\gamma(t)$ for the finial state $|\gamma,0;s\rangle$, and the finial-state probability $P_\gamma^{(s)}(t)$ which is given by $P_\gamma^{(s)}(t)=|c^{(s)}_\gamma(t)|^2$.

		\subsection{Absorption $A$ of Beam 2}
		
		Now we show our calculation for the results in Fig.~\ref{result2}(d), i.e., the 
		absorption $A(\theta)$ of beam 2, of left- and right-handed enantiomers. As mentioned in our maintext, we consider the system of Sec.~\ref{s3s}, with beam 1 and beam 2 being  a strong driving field and  a weak probe field, respectively. We further assume that spontaneous emission occurs from the state $|\gamma,0,s\rangle$ to the lower states $|\alpha,0,s\rangle$ ($s=L,R$) and $|\beta,\pm 1,s\rangle$, with all spontaneous emission rates, for simplicity, taken to be equal to $\kappa$. Thus, the quantum master equation of the molecule $s$ ($s=L, R$) is
		\begin{eqnarray} 
			\frac{d\hat{\rho}^{(s)}(t)}{dt} &=& -i[\hat{H}^{(s)},\hat{\rho}^{(s)}(t)]+\kappa\mathcal{L}[\overline{|\beta,+1;s\rangle}\overline{\langle\gamma,0;s|}]\hat{\rho}^{(s)}(t)+\kappa\mathcal{L}[\overline{|\beta,-1;s\rangle}\overline{\langle\gamma,0;s|}]\hat{\rho}^{(s)}(t) \nonumber\\
			&&+\kappa\mathcal{L}[|\alpha,0;s\rangle\overline{\langle\gamma,0;s|}]\hat{\rho}^{(s)}(t)+\kappa\mathcal{L}[|\alpha,0;s\rangle\overline{\langle\beta,+1;s|}]\hat{\rho}^{(s)}(t)+\kappa\mathcal{L}[|\alpha,0;s\rangle\overline{\langle\beta,-1;s|}]\hat{\rho}^{(s)}(t), \label{qme}
		\end{eqnarray}
		where $\hat{H}^{(s)}$ is given in Eq.~(\ref{hs2app}), and the Lindblad superoperators used in Eq.~(\ref{qme}) are defined by 
		\begin{equation}
			\mathcal{L}[|o_{1}\rangle\langle o_{2}|]\hat{\rho}^{(s)}(t)=[2|o_{1}\rangle\langle o_{2}|\hat{\rho}^{(s)}(t)|o_{2}\rangle\langle o_{1}|-|o_{2}\rangle\langle o_{2}|\hat{\rho}^{(s)}(t)-\hat{\rho}^{(s)}(t)|o_{2}\rangle\langle o_{2}|]/2.
		\end{equation} 
		By numerically solving Eq.~(\ref{qme}), we can obtain the steady-state density operator $\hat{\rho}^{(s)}_{\mathrm{ss}}$ of the molecule $s$, and the absorption of the beam 2 which is given by~\cite{Scully1997} 
		\begin{equation}
			A\propto\mathrm{Im}\left[\frac{\mathrm{Tr}[\hat{\rho}^{(s)}_{\mathrm{ss}}\hat{d}^{(s)}_{+}]}{(\Omega_{2}/2)}\right],
		\end{equation}
		with  $\hat{d}^{(s)}_{+}=e^{-i\theta_{\rm f}^{(s)}}\overline{|\beta,-1;s\rangle}\overline{\langle\gamma,0;s|}-e^{-i\theta}\overline{|\beta,+1;s\rangle}\overline{\langle\gamma,0;s|}$.

%


	\end{document}